\documentclass[conference]{IEEEtran}
\IEEEoverridecommandlockouts

\usepackage{cite}
\usepackage{amsmath,amssymb,amsfonts}
\usepackage{textcomp}
\usepackage{xcolor}
\usepackage{comment} 
\usepackage{float}  

\usepackage[noend]{algpseudocode}
\usepackage{amsmath}
\usepackage{array}
\usepackage{graphicx}
\usepackage{tabularx} 
 
\usepackage{caption}  
\usepackage{hyperref} 
\usepackage[numbers, square]{natbib}
\usepackage{xcolor}
\usepackage{amsmath}
\usepackage{algorithm}
\usepackage{algpseudocode}
\def\BibTeX{{\rm B\kern-.05em{\sc i\kern-.025em b}\kern-.08em
    T\kern-.1667em\lower.7ex\hbox{E}\kern-.125emX}}
\begin{document}

\title{{Two}-Agent DRL for Power Allocation and IRS Orientation  in Dynamic NOMA-based OWC Networks\\
}

\author{\IEEEauthorblockN{Ahrar N. Hamad, Ahmad Adnan Qidan, Taisir E.H. El-Gorashi and Jaafar M. H. Elmirghani}

}
\maketitle

%%%%%%%%%%%%%%%%%%%%%%%%%%%%%%%%%%%%%%%%%%%%%%%%%%%%%%%%%%%%%%%%%%%%%%%%
%%%%%%%%%%%%%%%%%%----------ABSTRACT------------%%%%%%%%%%%%%%%%%%%%%%%%
%%%%%%%%%%%%%%%%%%%%%%%%%%%%%%%%%%%%%%%%%%%%%%%%%%%%%%%%%%%%%%%%%%%%%%%%

\begin{abstract}

Intelligent reflecting surfaces (IRSs) technology has been considered a promising solution in visible light communication (VLC) systems due to its potential to overcome the line-of-sight (LoS) blockage issue and enhance coverage. Moreover, integrating IRS with a downlink non-orthogonal multiple access (NOMA) transmission technique for multi-users is a smart solution to achieve a high sum rate and improve system performance. In this paper, a dynamic IRS-assisted NOMA-VLC system is modeled, and an optimization problem is formulated to maximize sum energy efficiency (SEE) and fairness among multiple mobile users under power allocation and IRS mirror orientation constraints. Due to the non-convex nature of the optimization problem and the non-linearity of the constraints, conventional optimization methods are impractical for real-time solutions. Therefore, a two-agent deep reinforcement learning (DRL) algorithm is designed for optimizing power allocation and IRS orientation based on centralized training with decentralized execution to obtain fast and real-time solutions in dynamic environments. The results show the superior performance of the proposed DRL algorithm compared to standard DRL algorithms typically used for resource allocation in wireless communication. The results also show that the proposed DRL algorithm achieves higher performance compared to deployments without IRS and with randomly oriented IRS elements. 
\end{abstract} 


%%%%%%%%%%%%%%%%%%%%%%%%%%%%%%%%%%%%%%%%%%%%%%%%%%%%%%%%%%%%%%%%%%%%%%%%
%%%%%%%%%%%%%%%%%%---------INTRODUCTION------------%%%%%%%%%%%%%%%%%%%%%
%%%%%%%%%%%%%%%%%%%%%%%%%%%%%%%%%%%%%%%%%%%%%%%%%%%%%%%%%%%%%%%%%%%%%%%%

\section{Introduction}

Sixth-generation (6G) networks face the challenge of supporting ever-increasing user connectivity and bandwidth-intensive applications. To address this, researchers explore significant technologies such as optical wireless communications (OWCs) \cite{qidan2021towards}. More specifically, visible light communication (VLC) has emerged as a promising complementary technology to radio frequency (RF), which offers several advantages, including unlicensed spectrum utilization, immunity to RF interference, and enhanced spectral efficiency \cite{haas2015visible}. In indoor environments, VLC systems can significantly improve throughput and connectivity through the dual use of lighting infrastructure for both illumination and data transmission \cite{aletri2020optimum}. However, practical VLC deployments face critical challenges, particularly in maintaining reliable line-of-sight (LoS) connections in dynamic indoor environments where signal blockage and shadowing are common occurrences.

Intelligent reflecting surfaces (IRSs) technology, which reprovides an alternative path for signals to follow, has emerged as an important enabling technology for 6G networks \cite{basar2019wireless}. IRS is a surface that consists of several reflecting elements, such as metasurfaces or mirrors, which can be dynamically adjusted to control the reflection characteristics of incoming signals \cite{abdelhady2020visible}. The integration of IRS in  VLC systems has demonstrated improved data rate, enhanced coverage of the system, and the ability to overcome the LoS blockage challenge through steering non-line-of-sight (NLoS) reflection signals towards users \cite{aboagye2021intelligent}.

In multiuser VLC systems, employing non-orthogonal multiple access (NOMA) technique has the potential to significantly increase the VLC system throughput compared to conventional orthogonal multiple access (OMA) techniques \cite{kassahun2022performance}. In \cite{marshoud2018optical}, the integration of IRS and NOMA in VLC systems was investigated for enhancing spectral efficiency and user connectivity. However, this integration presents challenges in system optimization. For instance,  the joint optimization of power allocation coefficients and IRS configuration parameters becomes crucial in dynamic VLC systems \cite{OrientationsMU}. Moreover, conventional optimization algorithms, such as greedy algorithms, genetic algorithms, and iterative algorithms, are insufficient for real-time deployments where traffic demand and user distribution are subject to frequent updates \cite{abumarshoud2023intelligent}, \cite{sun2022joint}.

To tackle the aforementioned limitations, machine learning (ML) techniques have emerged as a promising solution to address the limitations of traditional algorithms in solving nonlinear and non-convex optimization problems \cite{jiang2016machine}. In \cite{arfaoui2021invoking}, the use of deep neural networks (DNNs) in VLC systems shows potential to estimate and predict key parameters for optimization. However, DNN models require extensive labeled data, which can be time-consuming and hard to obtain. Reinforcement learning (RL) is another ML technique considered as a promising tool to solve various optimization problems in wireless communication environments where the environment interacts with an agent and obtains rewards \cite{sutton2018reinforcement}, \cite{elgamal2021reinforcement}. Unlike supervised learning, RL agents navigate without explicit guidance, discovering policies through trial and error within the environment. For solving high-dimensional optimization problems with computational complexity, deep reinforcement learning (DRL) algorithms effectively address such challenges due to their model-free nature, which enables direct learning. In \cite{ciftler2021dqn}, \cite{zhao2019deep}, a deep Q-network (DQN) algorithm was used to solve complex resource allocation in heterogeneous wireless networks. However, DQN is a value-based method and inherently limited to discrete action spaces. To address this limitation, a deep deterministic policy gradient (DDPG) algorithm is developed to handle continuous state and action spaces by combining DQN principles with an actor-critic architecture \cite{Danya}, \cite{lillicrap2015continuous}. 

\subsection{Related Works and Motivation}
Recently, the integration of IRS into VLC systems has attracted attention from both academia and industry \cite{abdelhady2020visible}, \cite{aboagye2021intelligent}, \cite{sun2022joint}, \cite{sun2021intelligent}. 
In particular, analytical expressions were derived in  \cite{abdelhady2020visible} for IRS implementations in VLC systems using metasurfaces and mirror arrays. In \cite{aboagye2021intelligent}, an efficient sine-cosine algorithm was designed to determine the ideal IRS configuration and to create reliable NLoS links.  
 In \cite{sun2022joint}, the inherent blockage limitation of VLC with IRS integration was addressed by using a point-source model considering LoS and NLoS signals. 
 Furthermore, iterative algorithms called frozen variables and minorization-maximization were developed in \cite{sun2021intelligent} to solve a resource management optimization problem in an IRS-aided VLC system, aiming for enhanced spectral efficiency.

The application of NOMA in  IRS-assisted wireless networks has gained increasing interest, particularly in RF systems, for better use of resources. Existing studies have focused on optimizing various communication parameters, including power allocation, phase shift, and user pairing, to maximize spectral and energy efficiency \cite{wang2022sum}, \cite{hou2020reconfigurable}, \cite{cheng2021downlink}, \cite{ni2021resource}. In the context of OWC, only a few works in the literature have employed  NOMA for multi-user interference management in IRS-assisted VLC networks. For instance, in \cite{abumarshoud2022intelligent}, an adaptive-restart genetic algorithm was proposed to solve the optimization of IRS reflection coefficients, NOMA decoding order, and power allocation. 
Moreover, a low-complexity algorithm was introduced in \cite{liu2023sum} for optimal IRS passive beamforming in an IRS-assisted NOMA-based VLC system. However, several critical challenges remain unaddressed, including user mobility and time-varying traffic demand, where conventional algorithms for optimization are impractical, facing computational limitations in real-time.

For practical and instantaneous solutions, RL algorithms were explored as promising alternatives to consider for complex optimization problems in wireless networks \cite{gao2021machine}.  In \cite{ciftler2021dqn}, DQN was designed to optimize power allocation in hybrid RF/VLC networks to reach target data rates. In \cite{hammadi}, a deep Q-learning (DQL) framework was proposed for NOMA-VLC to optimize power allocation and LED transmission angles. Furthermore, a two-stage DQL resource management approach was designed to optimize spectral efficiency and fairness in indoor VLC systems.

It is worth pointing out that  DQL algorithms are effective for discrete action spaces, and they face significant constraints when handling continuous action spaces, particularly in high-dimensional scenarios. In OWC,  continuous control of the environment is essential due to optical channel fluctuations caused by various conditions such as user mobility and blockage. In \cite{Danya}, deep deterministic policy gradient (DDPG) algorithm was proposed to handle continuous state and action spaces in millimetre wave (mmWave) and VLC systems to optimize the secrecy capacity. Furthermore, in \cite{10511290}, DDPG was also proposed in a VLC system to optimize beamforming weights, IRS mirror optimization, and the phase shift to maximize the sum rate. Despite the ability of  DDPG to solve highly complex problems, it might face challenges in such scenarios, as it often becomes trapped in a local optimum, hindering it from learning effectively. A soft actor-critic (SAC) algorithm was proposed in \cite{zhang2024visible} with the goal of maximizing both the average reward and expected policy entropy. The problem was primarily designed for a single LED VLC system using OMA and for optimizing IRS orientation, time fraction assignment, and power allocation. However, the SAC algorithm demonstrated significant scalability constraints in larger-scale VLC networks with a high number of IRS elements, users, or LEDs. In \cite{zhang2025multi} the potential application of multi-agent deep reinforcement learning (MADRL) in 6G wireless networks was highlighted. Interestingly, the concept of MADRL can address the aforementioned limitations of employing RL for optimization in OWC networks, where it enables effective coordination among multiple agents in continuous action spaces, provides better scalability for complex environments,  and reduces the possibility of becoming trapped in local optimum through coordinated policy optimization.

\subsection{Contribution}
In this work, we model a dynamic IRS-assisted NOMA-VLC system that takes into account user mobility, blockage, and traffic demands. In contrast to the literature, two vital metrics, sum energy efficiency (SEE) and fairness, are optimized under {power allocation and IRS mirror orientation constraints}. A two-agent DRL algorithm is designed to handle the continuous action space of the formulated optimization problem to maintain scalability and adaptability in real-time.

The main contributions of this work are listed  as follows:

\begin{itemize}

\item An optimization problem is formulated to maximize the SEE and fairness while maintaining QoS requirements in a dynamic multi-user IRS-assisted NOMA VLC system. This optimization problem is defined as bi-fractional and non-convex due to the product of two functions,  SEE and fairness.

\item 
The non-convex optimization problem is reformulated as a Markov decision process (MDP), and a novel two-agent DRL algorithm is developed for solving the formulated MDP in real-time. Our DRL  is based on centralized training and decentralized execution for instantaneous solutions.

\item The performance of our proposed algorithm is validated through extensive numerical simulations in a real-time scenario where multiple users with time-varying traffic demands move at different velocities following the random waypoint (RWP) model. The results show that the proposed DRL algorithm achieves significantly high performance compared to standard DRL algorithms and other deployments without IRS and with randomly oriented IRS elements. 
The results also show that the proposed DRL effectively balances SEE maximization and user fairness.

\end{itemize}

The rest of this paper is organized as follows: Section \ref{sec:system} presents the system model for the IRS-aided NOMA-VLC network, including the LoS and IRS NLoS channel models, NOMA framework, power consumption, and user fairness. The optimization problem formulation for joint power allocation and IRS mirror orientation is presented in Section \ref{secIII}. Section \ref{sec:IV} introduces the proposed intelligent two-agent  DRL algorithm for solving the optimization problem. In Section \ref{sec:Results}, simulation results are presented. Finally, conclusions are provided in Section \ref{sec:con}.

%%%%%%%%%%%%%%%%%%%%%%%%%%%%%%%%%%%%%%%%%%%%%%%%%%%%%%%%%%%%%%%%%%%%%%%%
%%%%%%%%%%%%%%%%%%----------SYTSTEM MODEL------------%%%%%%%%%%%%%%%%%%%
%%%%%%%%%%%%%%%%%%%%%%%%%%%%%%%%%%%%%%%%%%%%%%%%%%%%%%%%%%%%%%%%%%%%%%%% 

\section{SYSTEM MODEL} \label{sec:system}
\begin{figure}
    \centering
    \includegraphics[width=0.9\linewidth]{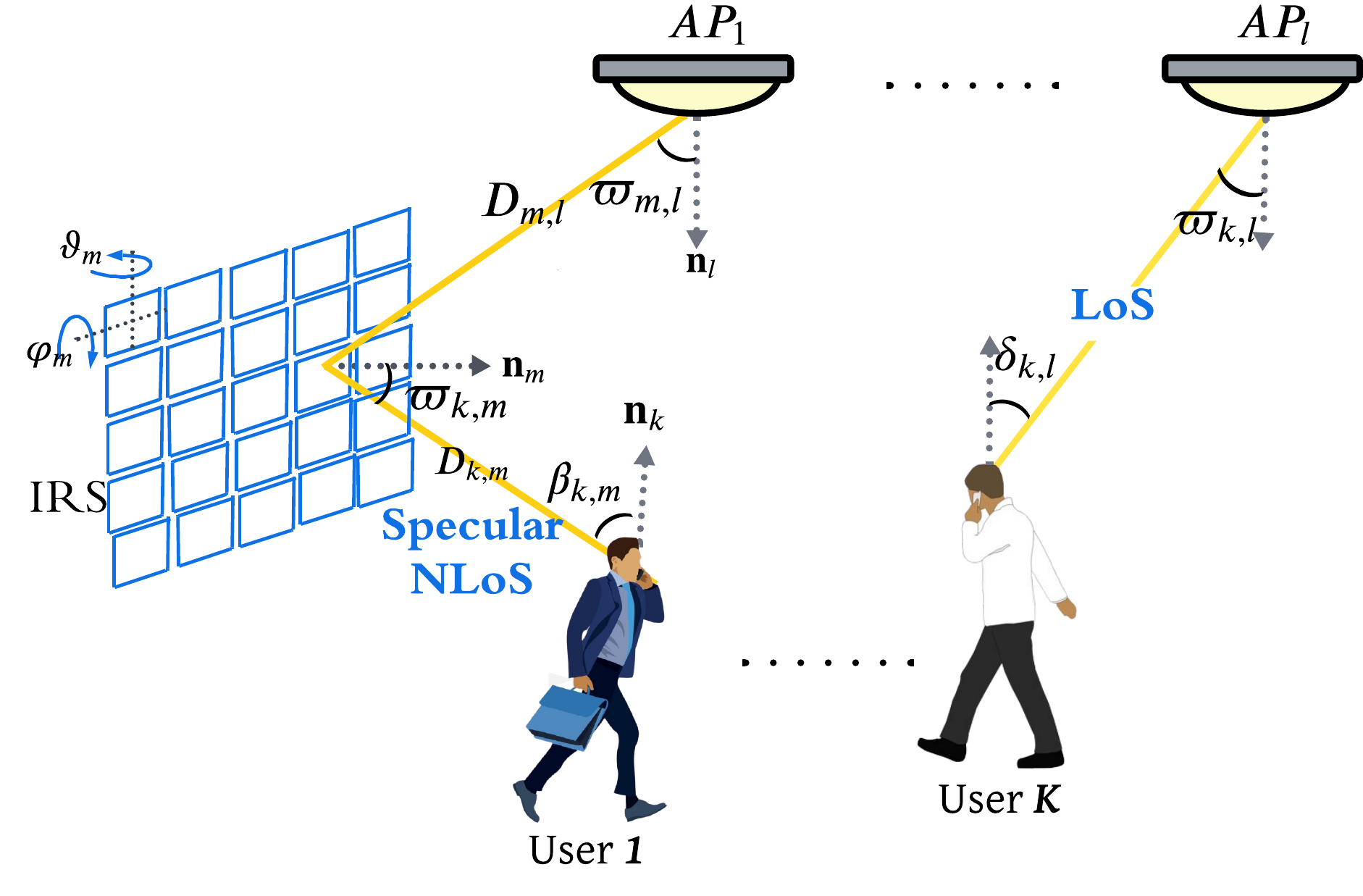}
    \caption{Indoor IRS-aided NOMA VLC system.} 
    \label{fig:system}
\end{figure}

We consider a downlink indoor IRS-aided NOMA VLC network in a room of dimensions  ($x_r, y_r, z_r$). The network consists of a set of multiple LED access points (APs), $\mathcal{L}$, $l = [1,\dots, L]$, placed on the ceiling serving a set of multiple mobile users, $\mathcal{K}$, $k=[1,\dots, K]$, and an IRS mirror array composed of a set of mirror elements, $\mathcal{M}$, $m=[1,\dots, M]$, as shown in Fig. \ref{fig:system}. 
Users are distributed on the receiving plane, and each user is equipped with an angle diversity receiver (ADR) to ensure connectivity, where each photodiode points to a specific direction determined by its azimuth $(Az)$ and elevation $(El)$ angles. For a dynamic network,  the users have time-varying traffic demands and move at different velocities following a RWP model \cite{bettstetter2004stochastic}. Moreover, 
the IRS mirror array deployed on one of the room walls consists of passive rotational mirrors to provide NLoS communication signals between the optical APs and users. Note that, each mirror $m$ has a rectangular shape with specific $w_m$ × $h_m$, representing the height and width, respectively. Also, a space $q_m$ is considered between the mirrors to ensure the interference-free characteristic between IRS reflected paths. As in \cite{abdelhady2020visible}, the IRS mirrors reflect optical signals to users via specular reflections, and each mirror can change its rotation as needed through its roll angle $\varphi$ and yaw angle $\vartheta$. In this work, we focus on the first-order reflection as in \cite{sun2022joint}, as higher-order reflections have a negligible impact on the network performance.

All optical APs and a WiFi AP are connected to a central unit (CU) that has essential information for managing the network and for executing the proposed algorithm. Moreover, joint transmission coordinated multi-point (JT-CoMP)  is considered across all the optical APs to work as one entity to ensure service continuity during user mobility as in \cite{arfaoui2022comp}.

%%%%%%%%%%%%%%%%%%%%%%%%%%%%%%%%%%%%%%%%%%%%%%%%%%%%%%%%%%%
%-------------------------- Figure 3 ------------------%

\subsection{The channel gain}
The direct LoS component of the optical channel is the primary component of the received power. The channel gain received by user \( k \in K\) from AP \( l \in \mathcal{L} \) considering the LoS signal is expressed as \cite{komine2004fundamental}

\begin{equation}
H_{k,l}^{\text{LoS}} = 
\begin{cases} 
\frac{(n+1) A_r \cos^n(\varpi_{k,l}) \cos(\delta_{k,l})}{2 \pi D_{k,l}^2}, & 0 \leq \delta_{k,l} \leq \psi_c \\
\quad \quad 0, & \delta_{k,l} > \psi_c
\end{cases}
\end{equation}

\noindent where \( n \) is the order of Lambertian emission, which is based on the half-power semi-angle of the LED \(\phi_{1/2}\) and can be calculated as \( n = -\ln(2)/\ln(\cos(\phi_{1/2})) \). Furthermore, \( A_r \) is the detector area,
\( \varpi_{(k,l)} \) is the he radiation angle between the normal to  AP \( l \) and the irradiance ray of user \( k \), \( \delta_{k,l} \) is the incident angle between the normal of the photodetector and the incident ray, and \( D_{k,l} \) is the distance between AP \( l \) and user \( k \).  To guarantee the signal of the direct LoS is detected by the receiver, the incidence angle \( \delta_{k,l} \) must be within a range from 0 to the acceptance semi-angle of the receiver concentrator (\( \psi_c \)), otherwise,  the LoS signal is not received. (i.e., $H_{k,l}^{\text{LoS}}$ = 0).

The NLoS signal is reflected by the mirror elements of the IRS array, assuming specular reflection. The received channel gain by user \( k \in \mathcal{K}\) from  mirror \( m \in  \mathcal{M}\) reflecting the signal from AP \( l \in \mathcal{L} \) is given by \cite{abdelhady2020visible} 

\begin{equation} \label{IRSCH}
H_{k,m,l}^{\text{IRS}} = 
\begin{cases} 
\frac{(n+1) \rho_m A_r dA_m \cos^n(\varpi_{m,l}) \cos(\beta_{m,l}) \cos(\varpi_{k,m}) \cos(\beta_{k,m})}{2\pi^2 (D_{m,l})^2 (D_{k,m})^2} \\
\quad \quad , 0 \leq \beta_{k,m} \leq \psi_c \\
\quad 0, ~~~~~~~\beta_{k,m} > \psi_c,
\end{cases}
\end{equation}

where \( \rho_m \) and \( dA_m \) are the reflection coefficient and area of mirror \( m \), respectively. Moreover,  \( \varpi_{m,l} \) is the radiation angle from AP \( l \) to mirror \( m \), $\beta_{m,l}$ is the incidence angle from AP $l$ to mirror $m$, $\varpi_{k,m}$ is the irradiance angle from mirror $m$ towards user $k$, and \( \beta_{k,m} \) is the incidence angle of the signal reflected from mirror \( m \) to user \( k \). Furthermore, \( D_{m,l} \) is the distance between AP \( l \) and mirror \( m \), and \( D_{k,m} \) is the distance between mirror \( m \) and user \( k \). In this work, the orientation of each IRS mirror can be adjusted through two angles, yaw  $\varphi_m$ and roll $\vartheta_m$,  where their impact on the user channel gain can be captured through  the cosine of $\varpi_{k,m}$ as follows

\begin{equation}
\label{IRSo}
\begin{aligned}
    \cos(\varpi_{k,m}) = & \bigg(\frac{x_m - x_k}{D_{k,m}}\bigg) \sin(\varphi_m) \cos(\vartheta_m) + \bigg(\frac{y_m - y_k}{D_{k,m}}\bigg)\\
    & \cos(\varphi_m) \cos(\vartheta_m)  + \bigg(\frac{z_m - z_k}{D_{k,m}} \bigg)\sin(\vartheta_m),
\end{aligned}
\end{equation}
\noindent where $(x_m, y_m, z_m)$ and  $(x_k, y_k, z_k)$ are the coordinate vectors for the locations of mirror $m$ and user $k$, respectively. Note that,   $\cos(\beta_{m,l})$ in \eqref{IRSCH} captures the impact of $\varphi_m$ and  $\vartheta_m$ angles on the channel gain between AP $l$ and mirror $m$, which can be derived easily similar to \eqref{IRSo}.

\subsection{NOMA}

The application of a power-domain NOMA scheme is considered to serve the users simultaneously over the same time or frequency slot by allocating different power levels based on the user channel gain \cite{chen2017performance}. The transmitted signal to $K $ users can be expressed as \cite{maraqa2023optical}
\begin{equation}
x(t) = \sum_{k=1}^K \sqrt{\alpha_k P_e} {s}_k(t) + I_{DC},
\end{equation}
where $\alpha_k$ is the power allocation coefficient for user $k$, $P_e$ is the transmitted power, ${s}_k(t)$ is the transmitted symbol to user $k$, $I_{DC}$ is the DC bias current to ensure non-negative signals. Note that, $P_e = \frac{P_{opt}}{q}$, where  $P_{opt}$ is  the optical power, and $q$ is the electrical-to-optical conversion ratio.
The sum of the power allocation coefficients for $K$ users is
\begin{equation}
     \sum_{k=1}^K \alpha_k = 1,
\end{equation}

At the other end of the communication link, the received  signal by user $k$  after removing the DC component can be expressed as

\begin{equation}
\begin{aligned}
y_k =  H_k  ({\sum_{k=1}^{K} {\sqrt{\alpha_k P_e}  \mathtt{s}_k(t)}} ) +n_k, 
\end{aligned}
\end{equation}

\noindent where $H_k= \sum_{l \in \mathcal{L}} H_{k,l}^{\text{LoS}} + \sum_{m \in \mathcal{M}} \sum_{l \in \mathcal{L}} H_{k,m,l}^{\text{IRS}}$is the channel gain for the combined LoS and IRS NLoS signals of user $k$, and $n_k \sim \mathcal{N}(0,\sigma^2)$ represents the additive white Gaussian noise (AWGN) including both thermal and shot noise components. 

Following NOMA principles \cite{chen2017performance}, users are sorted in order based on their combined channel gains as follows

\begin{equation}
|H_1| \leq |H_2| \leq \ldots \leq |H_K|.
\end{equation}
Accordingly, the power allocation coefficients are assigned inversely to channel strengths as follows
\begin{equation}
\alpha_1 \geq \alpha_2 \geq \ldots \geq \alpha_K.
\end{equation}

In line with the NOMA literature \cite{maraqa2023optical},  we consider a fixed decoding sequence according to the power values. The first step is to decode the strongest signal with the highest power while managing the remaining signals as noise. Then, the decoded signal is subtracted from the received signal. This process is repeated for the next strongest signal until all the signals are decoded. Therefore,
the signal-to-interference-plus-noise ratio (SINR), $\textit{$ \Gamma $}_k$, for user $k$ can be expressed as
\begin{equation}
\text{ $\textit{ $ \Gamma $}_k$} = 
\begin{cases}
\frac{(R_p H_k)^2 \alpha_k P_e}{(R_p H_k)^2 \sum_{j=k+1}^{K} \alpha_j P_e + BN_0}, & 1 \leq  k < K \\
\frac{(R_p H_k)^2 \alpha_k P_e}{BN_0}, & k = K
\end{cases}
\end{equation}

where $R_p$ represents the photodetector responsivity, $B$ is the system modulation bandwidth, and $N_0$ is the noise power spectral density.

In our IRS-assisted NOMA VLC system, we derive the achievable user rate considering the characteristics of intensity modulation/direct detection (IM/DD) and the dimmable VLC constraints \cite{wang2013tight}. Hence, the data rate for  user $k$  is given by\cite{wang2013tight}
\begin{equation}
R_k = B \log_2 \left(1 + \frac{\exp(1)}{2\pi} \Gamma_k \right),
\end{equation}
where $\frac{\exp(1)}{2\pi}$ accounts for the IM/DD constraints. Moreover, the system sum rate considering $K$ users  can be expressed as
\begin{equation}
R_T = \sum_{k=1}^K R_k.
\end{equation}

%%%%%%%%%%%%%%%%%%%%%%% -----  Energy Efficiency ----- %%%%%%%%%%%%%%%%%%%%%%%%%%%

\subsection{Power Consumption}
The system's total power consumption $P_{\text{total}}$ contains multiple components across the transmission from the optical AP to the receiver  \cite{maraqa2023optimized},\cite{aboagye2021energy}, which can be expressed as
\begin{equation}
P_{\text{total}} = P_{\textit{AP}} + P_{\textit{IRS}} + P_{\textit{Rec}}.
\end{equation}
\subsubsection{AP power}
The AP power consumption includes various components and can be calculated as
\begin{equation}
\begin{aligned}
P_{\textit{AP}} = & P_e + P_{\textit{circuit}}^{\text{TX}} + P_{\textit{{LED}}} + P_{{\textit{A}}} +
&P_{\textit{filter}}^{\text{TX}} + P_{\textit{{DAC}}},
\end{aligned}
\end{equation}
where $P_{\textit{circuit}}^{\text{TX}}=3250$mW is the transmitter circuit power, $P_{\textit{LED}}=2758$ mW is the LED driver power, $P_{\textit{A}}=280$ mW is the amplifier power consumption, $P_{\textit{filter}}^{\textit{TX}}=2.5$ mW is transmit filter power, and $P_{\textit{DAC}}=175$ mW is digital-to-analog converter power% and $P_e$ is the electrical transmit power.

\subsubsection{IRS power consumption}
The IRS power consumption scales linearly with the number of reflecting elements, which is given by
\begin{equation}
P_{\text{IRS}} = M P_{\text{element}},
\end{equation}
where $M$ denotes the number of IRS elements and $P_{\text{element}}=100$ mW is the power consumption for each mirror when steering.

\subsubsection{Receiver power}
The receiver power consumption can be calculated as
\begin{equation}
P_{\textit{Rec}} = P_{\textit{circuit}}^{\textit{RX}} + P_{\textit{filter}}^{\textit{RX}} + P_{\textit{TIA}} + P_{\textit{ADC}},
\end{equation}
where $P_{\textit{circuit}}^{\textit{RX}}=1.9$ mW is the receiver circuit power,
$P_{\textit{filter}}^{\textit{RX}}=2.5$ mW is the receive filter power,  $P_{\textit{TIA}}=2500$ mW is the trans-impedance amplifier power, and $P_{\textit{ADC}}=95$ mW is analog-to-digital converter power.

Considering the overall power consumption and the sum rate of the network, the SEE can be calculated as 
\begin{equation}
    {S_{EE}} = \frac{R_T}{P_{\text{total}}}.
\end{equation}

%%%%%%%%%%%%%%%% ---- Fairness ----- %%%%%%%%%%%%%%%%%
\subsection{Fairness}
To evaluate the user fairness of the network considered, we employ Jain's fairness index with a range of $[0,1]$, where 0 represents unfairness and 1 indicates perfect fairness among all users. The Jain fairness index is given by \cite{jain1984quantitative}
\begin{equation}
J = \frac{(\sum_{k=1}^K R_k)^2}{K \sum_{k=1}^K R_k^2}.
\end{equation}

%%%%%%%%%%%%%%%%%%%%%%%%%%%%%%%%%%%%%%%%%%%%%%%%%%%%%%%%%%%%%%%%%%%%%%%%
%%%%%%%%%%%%%----------PROBLEM FORMULATION------------%%%%%%%%%%%%%%%%%%
%%%%%%%%%%%%%%%%%%%%%%%%%%%%%%%%%%%%%%%%%%%%%%%%%%%%%%%%%%%%%%%%%%%%%%%%
\section{Problem Formulation and System Optimization}\label{secIII}
An optimization problem is formulated to maximize the SEE and ensure fairness under power allocation and IRS mirror orientation constraints. The objective function is given by

\begin{align} 
\label{P1}
\boldsymbol{\mathrm{P_1:}} \quad & \underset{\boldsymbol{\alpha}, \boldsymbol{\varphi}, \boldsymbol{\vartheta}}{\text{maximize}} \quad J \cdot S_{EE},
\end{align}
\text{subject to:}
\begin{align}
& ~~~~~~~R_k \geq R_{\text{min}}, \quad \forall k \in \{1, \ldots, K\}, \tag{18.a} \label{18.a}\\ 
& ~~~~~~~ P_e \leq P_{\text{max}}, \tag{18.b}\label{18.b} \\ 
&~~~~~~~\sum_{k=1}^K \alpha_k = 1, \tag{18.c} \label{18.c}\\
& ~~~-~\frac{\pi}{2} \leq {\varphi_m} \leq \frac{\pi}{2}, \quad \forall m \in \{1, \ldots, M\}, \tag{18.d}  \label{18.d}\\ 
&~~~~ -\frac{\varpi}{2} \leq {\vartheta_m} \leq \frac{\pi}{2}, \quad \forall m \in \{1, \ldots, M\}, \tag{18.e} \label{18.e}
\end{align}

\noindent where constraint \eqref{18.a} ensures the minimum QoS requirements for each user, while constraint \eqref{18.b} ensures the sum of the power allocated to the users does not exceed the maximum power budget. Moreover,  constraint \eqref{18.c} is defined to ensure that the total transmit power equals to $P_e$, and constraints \eqref{18.d} and \eqref{18.e} are limits for IRS mirror orientation.

The optimization problem \textbf{P1} is highly non-convex under nonlinear constraints. One of the ways to solve the optimization problem is to use successive convex approximation (SCA) \cite{tahira2019optimization}. In this approach,  the first-order Taylor approximation is applied to transform the objective function into a convex form and to handle the IRS mirror orientation, while a first-order approximation is applied for the power constraints. By iterating over these approximations, a sub-optimal solution can be provided with polynomial complexity of  $\mathcal{O}(K^{3}+M^{3})$ at each iteration. Therefore, advanced approaches are required for fast solutions in a dynamic real-time environment with frequent variations in user mobility and traffic demands.

Multi-agent is particularly well-suited for our problem due to its ability to handle complex non-convex optimizations and make real-time decisions in a dynamic environment \cite{zhang2021multi}.
In the next section, we propose a two-agent DRL to handle the formulated optimization problem in \eqref{P1}. The first agent is to determine the power allocation coefficients for the users. The second agent focuses on steering the mirrors dynamically towards the users, which is crucial, especially for mobile users experiencing poor SNR due to their speed.

%%%%%%%%%%%%%%%%%%%%%%%%%%%%%%%%%%%%%%%%%%%%%%%%%%%%%%%%%%%%%%%%%%%
%%%%%%%%%%%%%%%%%%%%%----------MADDPG Algorithm------------%%%%%%%%%%%%%%%%%%%%%%%%%%%%%%%%%%%%%%%%%%%%%%%%%%%%%%%%%%%%%%%%%%%%

\begin{figure}
    \centering
    \includegraphics[width=\linewidth]{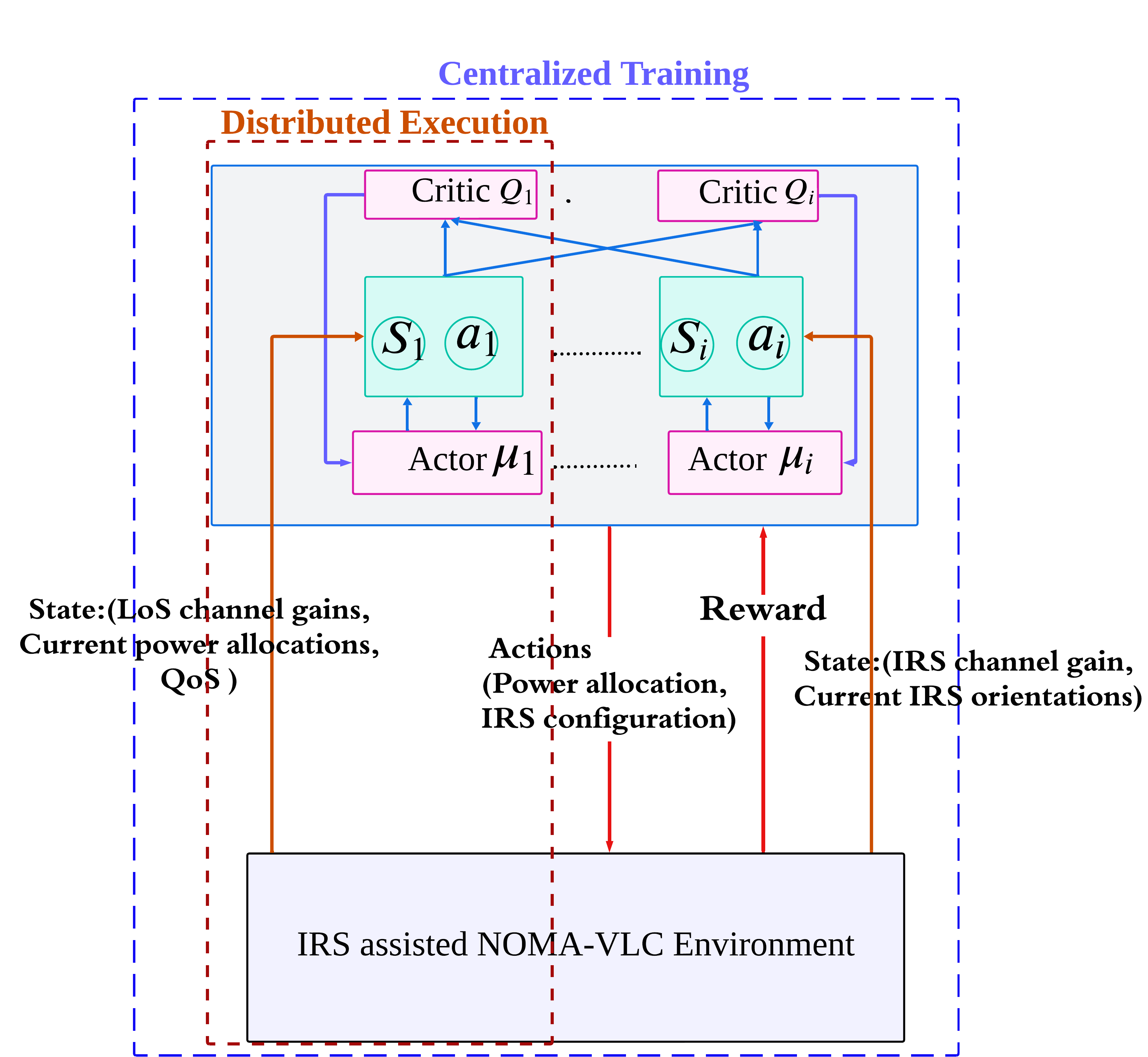}
    \caption{Two-agent DRL based on centralized training and decentralized execution framework.}
    \label{fig:MADDPG}
\end{figure}

\section{Intelligent Solution Using A DRL Algorithm} \label{sec:IV}

A two-agent DRL algorithm is proposed in this section, starting with a reformulation of the optimization problem as a MDP model. We then provide a detailed implementation of the proposed algorithm. 

%%%%%%%%%%%%%%%%%%%%%%%%%%%%%%%%%%%%%%%%%%%%%%%%%%%%%%%%%%%%%%%%%%%%%%%
\subsection{MDP Formulation}\label{subsec:A}
%%%%%%%%%%%%%%%%%%%%%%%%%%%%%%%%%%%%%%%%%%%%%%%%%%%%%%%%%%%%%%%%%%%%%%%

The optimization problem \textbf{P1} is reformulated as a MDP model defined by the tuple $\langle\mathcal{S}, \mathcal{A}, \mathcal{P}, \mathcal{R},$ $\mathcal{\gamma}\rangle$, where  $\mathcal{S}$ is the state space, $\mathcal{A}$ is the action space of the agents, $\mathcal{R}$ is the reward representing the system objectives in \eqref{P1}, $\mathcal{P}$ is the probability of the state transition, and $\gamma$ is the discount factor for future rewards. 

For the modeled IRS-assisted NOMA VLC environment, we employ two agents,  $AG_l$ and $AG_m$, where during the training, the agents exchange information to incorporate the global environment context. However, during execution, the two agents act in parallel and independently. This decentralized approach is essential in real-time, allowing each agent to make independent decisions based on limited information. For instance, at a time slot $t$, each agent observes its current state $s_t \in S$ and selects an action $a_t \in \mathcal{A}$ based on its individual policy. Each agent has a local observation $o_i$, $i \in \{l,m\}$, which is part of the global state. The environment then transitions to the next state $s_{t+1}$ according to the transition probability $\mathcal{P}(s_{t+1} | s_t, a_t^1, \dots, a_t^{\mathcal{N}_a})$, where $\mathcal{N}_a$ represents the total number of agents. The goal of the agent is to maximize its cumulative reward according to our optimization problem \textbf{P1}. The state space, the action space, and the reward are designed as follows.

%%%%%%%%%%%%%%%%%%%%%%%%%%%%%%%%%%%%%%%%%%%%%%%%%%%%%%%%%%%%%%%%%%%%%%%%%%%%%
The state space $\mathcal{S}$ consists of a set of all possible states $s_t \in \mathcal{S}$. At time step $t$, each agent receives its own observation $o_{i,t}$ from the environment. The observation $o_{l,t}$ of the $AG_l$ agent consists of the SNRs of the $K$ users,  $\boldsymbol{\Gamma}_{k}\in \mathbb{R}^{1\times K}$, minimum data rate of each user to ensure high QoS, $\boldsymbol{R}_{min,k} \in \mathbb{R}^K$, and the current IRS rotation angles, $\boldsymbol{\varphi}_{m,t} \in [-\pi]/2,\pi/2]^M$ and $\boldsymbol{\vartheta}_{m,t} \in [-\pi/2,\pi/2]^M$. On the other hand, 
the observation $o_{m,t}$ of the $AG_m$ agent consists of the SNRs of the $K$ users $\boldsymbol{\Gamma}_{k} \in \mathbb{R}^{1\times K}$, current power allocation coefficients $\boldsymbol{\alpha}_{k} \in [0,1]^{1\times K}$, and minimum data rate of each user.

The action space $\mathcal{A}$ contains all the actions, which are the outputs of the actor networks of the agents at each time slot $t$. Each agent chooses an action $a_t$ $\in$ $\mathcal{A}$ based on its observed state $o_i$ and must satisfy the constraints in \eqref{P1}.

The first agent chooses an action $a_{l,t}$ $\in$ $\mathcal{A}$ to determine the power allocated to each user, ${\alpha}_{k,t} $. The selected action must satisfy the NOMA constraints in \ref{18.b} and \ref{18.c},  and can given by
\begin{equation}
a_{l,t} = \begin{bmatrix} \alpha_k | k \in \mathcal{K}\
\end{bmatrix},
\end{equation}
The second agent takes action $a_{m,t}$ to determine the IRS angles  $ \boldsymbol{\varphi}_{m,t}$, and $\boldsymbol{\vartheta}_{m,t}$. This allows the IRS to steer the reflected signals towards the users and under the IRS constraints in \ref{18.d} and \ref{18.e}. This action is  given by
\begin{equation}
a_{m,t} =  \begin{bmatrix}\varphi_m, \vartheta_m | m \in \mathcal{M}\ \end{bmatrix},
\end{equation}

Now, the reward must reflect the objective function of the optimization problem \textbf{P1}. Each agent receives a reward $r_i \in \mathcal{R}$, $i \in \{AG_l, AG_m\}$,  from the environment determined based on the current state $s_t$, the action $a_t$ taken at time step $t$, and the next state $s_{t+1}$ received from the environment. Note that, the agent must only receive a reward if the selected actions satisfy all the constraints to maximize the long-term expected return. Hence, the global reward $\mathcal{R}$ is computed as the sum of the individual agent rewards as

\begin{equation}
\mathcal{R}_t = \sum_{t=0}^{T} \gamma_t r_t(s_i,a_i),
%r_t = \sum_{t=0}^{T} \gamma_t SEE_{k,t},
\end{equation}

\noindent where $\gamma \in [0,1]$ is the discount factor to discount the value of future rewards. To ensure the system constraints are within the reward design, we introduce two penalty terms. The first penalty term is
$\varrho_{1,t}$ = $\sum_{k=1}^K [ R_{k,t}<R_{min,t} ]$, which addresses the minimum QoS requirement constraint. This term also counts the total number of users whose achievable rates fall below the minimum required rate $R_{min}$ at time step $t$. The second penalty term is $\varrho_{2,t}$ = $[ P_{e,t}  > P_{\max,t}]$, which is formulated to force the power constraint when the power consumption exceeds the maximum power constraint.   Therefore, the reward function can be defined as 
\begin{equation}
r_t = \sum_{k=1}^K SEE_{k,t} +\sum_{k=1}^K J_{k,t} -\lambda_1 \varrho_{1,t} - \lambda_2 \varrho_{2,t}.
\end{equation}

\noindent where $\lambda_1$ and $\lambda_2$ are  trade-off coefficients to balance the penalties. Note that, the power allocation constraints $\sum_{k=1}^K \alpha_k = 1$ and IRS angle constraints $\omega_m, \phi_m \in [-\pi/2, \pi/2]$, are enforced through action space design. The penalties punish the reward when constraints are violated, helping to guide the learning process toward feasible solutions and satisfy the constraints. This helps the agents to maximize the reward while minimizing the penalties.

It is worth mentioning that the formulated MDP  can be solved using standard and advanced DRL algorithms such as DDPG. However, this DRL algorithm treats the environment as unstable and relies on noise-based exploration \cite{dankwa2019twin}, which is not efficient in our highly dynamic network where users might travel at potentially high velocities with time-varying traffic demands.

\subsection{Two-Agent DRL Policy}

Before going into the proposed algorithm for our IRS-assisted NOMA-VLC system, we first present the policy methods designed for continuous control problems \cite{jia2022policy}.

\begin{itemize}
    \item \textit{Stochastic Policy Gradient:} In policy gradient methods, the agent’s behavior is characterized by a parameterized policy, $\pi_\theta(a|s)$, which defines a probability distribution over actions given the current state. The objective is to maximize the expected cumulative discounted reward given by

\begin{equation}
J(\theta) = \mathbb{E}_{s \sim \rho^\pi, a \sim \pi_\theta} \left[\sum_{t=0}^T \gamma^t r_t\right],
\end{equation}

where $\rho^\pi$ represents the probability of the state distribution under policy \( \pi \). 
The policy gradient theorem provides the fundamental mechanism for policy optimization given by
\begin{equation}
\nabla_\theta J(\theta) = \mathbb{E}_{s \sim \rho^\pi, a \sim \pi_\theta} \left[\nabla_\theta \log \pi_\theta(a|s) Q^\pi(s,a)\right],
\end{equation}

where  $\pi$ denotes the stochastic policy, \( \pi_\theta(a \mid s) \) is the policy mapping state \( s \) to action \( a \), parameterized by \( \theta \), \
\( Q^\pi(s, a) \) is the action-value function, representing the expected cumulative reward for state \( s \) and action \( a \), and following policy \( \pi \), and \( \nabla_\theta \) is the gradient for the policy parameters \( \theta \). 
In this framework, the policy $\pi_\theta$ shows stochastic behavior, where actions are sampled according to the conditional probability density $\pi_\theta(a|s)$. However, a significant challenge arises when dealing with high-dimensional action spaces, as the policy gradient algorithm requires extensive sampling, leading to substantial computational overhead.

    \item \textit{Deterministic Policy Gradient:}
For our indoor environment with continuous states and actions spaces, the stochastic policy gradient faces challenges in high-dimensional action spaces due to extensive sampling requirements. Hence, the deterministic policy gradient (DPG) can address this by learning a deterministic policy $\mu_\theta: \mathcal{S} \rightarrow \mathcal{A}$, which directly maps state $s$ to action $a$ and can be expressed as \cite{9576103}

\begin{equation}
\nabla_\theta J(\theta) = \mathbb{E}_{s \sim \rho^\mu} \left[\nabla_\theta \mu_\theta(s) \nabla_a Q^\mu(s,a)|_{a=\mu_\theta(s)}\right],
\end{equation}

\noindent where $\rho^\mu$ is the state distribution under the deterministic policy $\mu_\theta$.
Note that, our system benefits from the DPG because it produces several advantages, such as computational efficiency and precision control.

\end{itemize}
This theoretical framework provides the foundation for our implementation design, enabling efficient joint optimization of power allocation and IRS orientation to maximize SEE and fairness.

We employ the DPG policy for efficient optimization through state-space integration. The first agent can have a deterministic policy for the power allocated to $K$ users, and the second agent can have a deterministic policy for $M$ IRS mirror orientations. Both agents, $AG_l$ and $AG_m$,  can learn optimal policies while considering each other's actions during training. The expected return gradient formulation for each agent $AG_i$, $i \in \{l, m\}$, is given by

\noindent
\begin{equation}
\begin{aligned}
\nabla_{\theta^i} J (\mu_{\theta^i}) &= \mathbb{E}_{o,a \sim \mathcal{D}} 
\Big[ \nabla_{\theta^i} \mu_{\theta^i}(o^i) \\ % (s^i | a^i)
&\quad \times \nabla_{a^i} Q_i^\mu (\textbf{o}, a_l, a_m) 
\Big|_{a^i=\mu_{\theta^i}(o^i)} \Big].
\end{aligned}
\label{eq:policy_MADDPG}
\end{equation}

\noindent where $\mu_{\theta^i}$ is deterministic policy of the agent, $\theta^{i}$ are the policy parameters, $s_i$ is the local observation for agent $i$,
$Q_i^\mu$ is the centralized action-value function, and $\mathcal{D}$ is the replay buffer containing experience tuples. Note that, $\textbf{o} =(o_l, o_m )$ is the observation of two agents along with information.

%------------ Figure 2---------------------%
\begin{figure}
    \centering
    \includegraphics[width=1\linewidth]{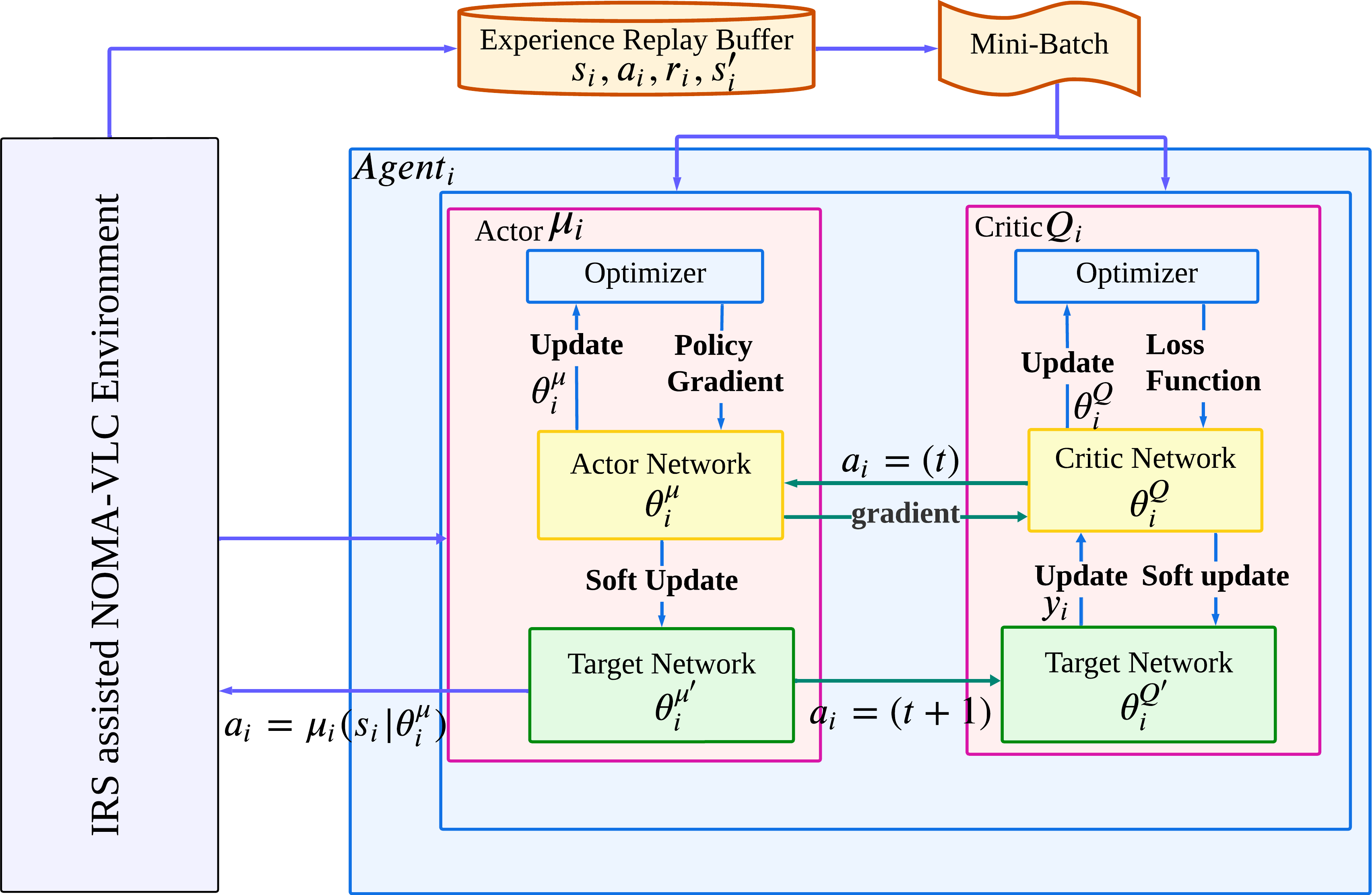}
    \caption{Training diagram of an agent.} %arc
    \label{fig:Agent}
\end{figure}
\subsection{Two-Agent DRL: Learning and  Execution } \label{CTDE}

We consider centralized learning and decentralized execution for the proposed two-agent DRL as shown in Fig. \ref{fig:MADDPG}. This mechanism enables the agents, $AG_l$ and $AG_m$, to learn policies that account for the joint impact of NOMA power allocation and IRS orientation on system performance.

In the proposed algorithm and as shown in Fig. \ref{fig:Agent}, each agent $AG_i$, $i \in \{l,m\}$, has an actor network $\mu_{\theta_i}$ and a critic network $Q^{\mu}_{\phi_i}$ as shown in Fig. \ref{fig:Agent}. In this context, the implementation of centralized learning and decentralized execution  can be described as follows

\subsubsection{Centralized Training Phase} 
During offline training, for each agent $AG_i$, $i \in \{l,m\}$, the actor-network $\mu_{\theta_i}$ maps local observations $o_i$ to actions $a_i$ according to the inputted mini-batch $N_b$ of transitions $j$, while its critic $Q^{\mu}_{\phi_i}$ evaluates action-value functions $Q$ using global information (i.e., the global state $s$ or all agent observations $\textbf{\textit{o}}$, and all agents’ actions ($a_l, a_m$) to ease the training. Furthermore, using the whole information allows each agent to
learn its state-action value function separately and be aware of other agents actions. For the power allocation agent, the critic $Q^{\mu}_{l}$ evaluates power coefficients, $\alpha_k$ for each user, considering current IRS orientation angles, while the critic of the IRS agent $Q^{\mu}_{m}$ evaluates the IRS mirror angles,  $\varphi$ and $\vartheta$ for each mirror element, considering current power allocation coefficients, ensuring a collective impact on overall system performance.

For enhanced training stability and efficiency, experiences are stored in replay buffer $\mathcal{D}$, which is used to save and reuse past training samples $\mathcal{D} $= $\{(o_j, a_j, r_j, o'_{j})\}$. Note that, the oldest samples must be removed when the replay buffer is full. Moreover,

The centralized critic network is updated by minimizing the loss function given by
\begin{equation} \label{eq:loss}
\mathcal{L}({\phi_i}) = \mathbb{E}_{{o}, a,r,{o'}\sim\mathcal{D}}[({Q}_i^{{\mu}}\boldsymbol{o},a_l,a_m ;\phi_i) - y_i)^2],
\end{equation}

\noindent where ${Q}_i^{{\mu}}({o},a_l,a_m)$ is the predicted Q-value from the current critic-network and the target value $y_i$ is calculated as  

\begin{equation} \label{eq:target}
y_i = r_i + \gamma  {Q}_i^{{\mu'}}(\textbf{o}',a'_l,a'_m; \phi'_i))|_{a'_i=\boldsymbol{\mu}'_i(o'_i)},
\end{equation}
\noindent where $r_i$ is the immediate reward for agent $AG_i$, and  $o'$ is the next global state. Moreover,  $a'_l$ and $a'_m$ are the next actions for the two  agents,   $\mu'_i$ is the target policy network for agent $AG_i$, and ${Q}_i^{{\mu'}}(\textbf{o}',a'_l,a'_m;\phi'_i)$ is the target critic network with parameters $\phi'_i$.

Furthermore, the actor-network $\mu_{\theta_i}$ in each agent $AG_i$ is updated using mini-batch $\mathcal{N}_b$ sampling from the buffer of the agent, which is given by
\noindent
\begin{align} \label{eq:Actorsampling}
\nabla_{\theta^{\mu_i}} J(\mu_{\theta_i}) \approx & \frac{1}{\mathcal{N}_b} \sum_{j=1}^{N_b}[ \nabla_{\theta^{\mu_i}} \mu_i(\boldsymbol{o}_i^j) \nonumber \\ 
& \times \nabla_{a_i} Q_i(o^j,a_l^j,a^j_m))|_{a_i^j=\boldsymbol{\mu}_i(o_i^j)}],
\end{align}

 The process above ensures efficient optimization in continuous action spaces, stable learning through experience replay, coordinated multi-agent policy updates, and constraint satisfaction for both agents. It is worth mentioning that during training, adding Gaussian noise  $\mathcal{N}_t \sim \mathcal{N}(0,\sigma^2)$  to the actor policy helps ensure balance between exploration and exploitation.

\subsubsection{Decentralized Execution Phase} 

During execution, each agent $AG_i$, $i \in \{l,m\}$,  uses its local observation $o_i$ by the actor-network, and selects action independently using its policy,  where the policy $\mu_{\theta_{i}}$ maps the observation directly to action space. Each agent selects its action $a_i$ for a given observation $o_i$ by

\noindent
\begin{equation} \label{eq:decentralize}
a_i = \mu_{\theta_i} (o_i) : O_i \rightarrow \mathcal{A}_i, \quad i \in \{AG_l,AG_m\}, 
\end{equation}
Note that, the parameters $\theta_i$ are fixed during execution, this means there are no gradients, loss functions, or replay buffer sampling like in training. The target actor network $\mu'_{\theta'_i}$ and the target critic networks $Q'^{\mu}_{\phi'_i}$ are updated their parameters using the soft update approach, which enables smoother learning dynamics and enhanced training stability. The target networks parameters are updated once per episode as

\begin{equation} \label{eq:updated}
\begin{aligned}
 \theta^{\mu'}_i  &\leftarrow \tau_o\theta_i ^{\mu} + (1-\tau_o)\theta_i ^{\mu'} \\
\phi'_i  &\leftarrow \tau_o\phi_i + (1-\tau_o)\phi'_i
\end{aligned}.
\end{equation}

\noindent where $\tau_o \ll$ 1 donates the weight decay factor.

%%%%%%%%%%%%%%%%%%%%%%%%%%%%%%%%%%%%%%%%%%%%%%%%%%%%%%%%%%%%%%%%%%%%%%%%%
%--------- MADDPG ALGORITHM ---------%
%%%%%%%%%%%%%%%%%%%%%%%%%%%%%%%%%%%%%%%%%%%%%%%%%%%%%%%%%%%%%%%%%%%%%%%%%
\begin{algorithm}[t]
\caption{The Proposed DRL for Joint Power Allocation and IRS Mirror Orientation}\label{MA_alg}

\textbf{Initialize:}  

\begin{algorithmic}[1] 
\State LED power allocation agent's actor $\mu_{LED}$ and critic $Q^{\mu}_{LED}$ with parameters $\theta^{\mu_{LED}}$ and $\phi_{LED}$
\State  IRS agent's actor $\mu_{IRS}$ and critic $Q^{\mu}_{IRS}$ with respective parameters $\theta^{\mu_{IRS}}$ and $\phi_{IRS}$
\State Target networks $\mu'_{LED}$, $Q'^{\mu}_{LED}$, $\mu'_{IRS}$, $Q'^{\mu}_{IRS}$ with parameters $\theta^{\mu'_{LED}}$, $\phi'_{LED}$, $\theta^{\mu'_{IRS}}$, $\phi'_{IRS}$
\State Experience replay buffer $\mathcal{D}$
\ \textbf{Training}
\For{episode = 1 to ${N_{ep}}$}
    \State Observe initial observations $o_l$ and $o_m$ and set the reward =0
    \For{t = 1 to \textit{Timestep}}
        \State select action for each agent using Eq. \ref{eq:decentralize}
        \State Execute actions $a_t=\{{a_{l,t},a_{m,t}\}}$, 
        calculate the reward $r_t$ , and  transform to new observation $o'$ for each agent. 
\For{each agent $AG_i$} 
    \If {the number of transition $< \mathcal{N}_b$}
        \State Store the transition tuple in $\mathcal{D}$
        \State Replace the earliest saved transitions in the buffer with $\{o_t, a_t, r_t, o'_t\}$
        \State Randomly select a mini-batch of transitions $\{o^j_t, a^j_t, r^j_t, o'^j_t \}$ from the replay buffer $\mathcal{D}$
        
        \State Update the agent using Eq. \ref{eq:target}
        \State Update the critic training network by minimizing the loss in the Eq. \ref{eq:loss}
        \State Update the actor training network using policy gradient Eq. \ref{eq:Actorsampling}
        \State Update actor and critic target networks using Eq. \ref{eq:updated}
    \EndIf
    \State $s \leftarrow s'$
\EndFor
\EndFor
\EndFor
\end{algorithmic}
\end{algorithm}

%-------------------------------------------------%
\subsection{Overall Algorithm Description} 

The proposed DRL  algorithm is described in Algorithm \ref{MA_alg}. The training process starts by initializing crucial training parameters of two agents, the power allocation agent $AG_l$ and the IRS mirror orientation agent $AG_m$,  along with the replay buffer $\mathcal{D}$ as in steps 1-4. In step 5, the training process is organized into $N_{ep}$ episodes, where each episode contains $T$ time steps. At each time step $t$, the power allocation and  IRS mirror orientation agents observe their states described in \ref{subsec:A}  as in step 6 while initializing the rewards. Based on these observations, each agent generates continuous actions through their respective actor networks with exploration noise $N_t$. Note that, the $AG_l$ agent determines  the power allocated  to each user, $\alpha_k$,  while considering constraints \eqref{18.b} and \eqref{18.c},  and the $AG_m$ agent determines optimum mirror orientation  angles, $\varphi_m$ and $ \vartheta_m$,  under constraints \eqref{18.d} and \eqref{18.e}. Therefore, each agent selects an action $a_{i,t}$, $i \in \{l,m\}$, and executes this action independently, as in steps 8-9 to learn their rewards, $r_{i,t}$.  In steps 10-14, the experience replay mechanism is controlled, as it is important for stable learning in multi-agent environments. At first, the current transition tuple consisting of $\{o_t,a_t,r_t,o'_t\}$ is saved to the replay buffer $\mathcal{D}$. When the buffer is full, the algorithm uses the first in first out policy to erase the oldest data from the buffer and replace it with the newest data to keep the size of the buffer constant and also to keep the recent experiences. To provide diverse learning samples, the algorithm randomly selects a mini-batch of transitions $j$ from the buffer $\mathcal{N}_b$. Each transition in this mini-batch from different time steps in the past can be used to update the agents' neural networks diversely, stabilizing the training process. In steps 15-18, the algorithm executes the core two-agent DRL training updates for both the actor and critic architectures. Step 16 focuses on updating the critic networks by minimizing the loss function, while step 17 updates the actor networks using the policy gradient method. Finally, step 18 updates the target networks of the actors and critics. After centralized training, each agent executes actions independently, i.e., decentralized execution,  based on only local observations.

\subsection{Computational Complexity}

The computational complexity of the proposed DRL algorithm can be analyzed through both actor and critic networks for both agents. Interestingly, the actor-network takes the state as an input and displays the action as an output, while the critic network takes the state and action as an input and displays a Q value as an output. For the agent $AG_l$, the state dimension equals to $2K+2M$, and its action dimension equals to $K$. Therefore, the complexity of the actor-network is $\mathcal{O}((2K+2M) \times K)$, and the complexity of the critic network is  $\mathcal{O}((2K+2M)+K)$. Similarly, for the agent $AG_m$,  the state dimension equals to $3K$, and its action dimension equals $2M$. Therefore, the complexity of the actor-network of this agent is $ \mathcal{O}(3K \times 2M)$, and the complexity of its critic network is $\mathcal{O}(3K+ 2M)$.

%%%%%%%%%%%%%%%%%%%%%%%%%%%%%%%%%%%%%%%%%%%%%%%%%%%%%%%%%%%%%%%%%%%%%%%%%%
\section{Simulation and Performance Evaluation}\label{sec:Results}
In this section, we evaluate the performance of the proposed algorithm for solving the formulated optimization problem in an indoor IRS-assisted NOMA-VLC environment.

\subsection{System Configuration}

We consider an indoor room with a size of $(5m \times 5m \times 3m)$ with four optical APs transmitters $L = 4$ placed on the ceiling and five active users, $K$ = 5, of different heights with dynamic traffic demand. A minimum data rate of 1 Mbps, $R_{min}= 1$ Mbps,  unless specified otherwise. In our system, each user moves in the room following RWP at velocities $v_{k} \in [0, 2]$ m/s chosen corresponding to typical indoor mobile user speeds.

Moreover, an IRS is mounted on one wall and consists of $7 \times 7$ mirror elements with the centre point (0 m, 2.5 m, 1.5 m). Each mirror element size is $25cm \times 15 cm$. Each mirror element is spaced 10 cm apart \cite{sun2022joint}. Other simulation parameters are shown in Table \ref{tab:params}.

%---------------------- TABLE 1---------------%
\begin{table}[t]
\caption{Simulation Parameters.}
\label{tab:params}
\begin{center}
\begin{tabular}{|l|c|}
\hline
\textbf{Parameter} & \textbf{Value} \\
\hline
Room dimensions & $5m \times 5m \times 3m$ \\
%LED transmitter height & 3m \\
Number of LED transmitters & 4 \\
Number of IRS elements & 49 $(7 \times7$ array) \\
Number of users & 8 \\
LED semi-angle at half power & 60° \\
%Receiver FOV & 90° \\
Detector area & 1 cm² \\
Optical filter gain & 1.0 \\
Refractive index & 1.5 \\
Photodetector responsivity & 0.5 A/W \\
Noise power spectral density & $10^{-21}$ $A^2/Hz$ \\
System bandwidth & 20 MHz \\
%Minimum required data rate & 10 Mbps \\
\hline
\end{tabular}
\end{center}
\end{table}
%-----------------------------------------------%

\subsection{Training Configuration}

The simulation is implemented using Python 3.12 and PyTorch 2.0. The actor and critic networks are fully connected and updated using an adaptive moment estimation (ADAM) optimizer because of their robust adaptation to non-stationary objectives. Moreover, the actor-network of the LED power allocation agent $AG_l$ consists of an input layer with $3Z_i$ neurons corresponding to SNRs, current IRS mirror angles, and QoS values. This input layer is followed by two hidden layers, each containing $128$ neurons.  The rectified linear unit $ReLU (.)$ activation function $f_{ReLU}(x) = \max(0,x)$ is employed in the hidden layers in both actor networks. The output layer comprises $Z_o$ neurons with $Softmax$ activation to ensure valid power allocation coefficients. Similarly, the actor-network of the IRS orientation agent $AG_m$ consists of an input layer of $Z_i$ neurons. The input layer of this agent corresponds to SNRs, QoS values, and current power allocation. This input layer is followed by two hidden layers, each with $64$ neurons, passing to $Z_o$ output neurons with an activation function $f_{Tanh}(x) = \frac{e^x - e^{-x}}{e^x + e^{-x}}$, $Tanh (.)$,  for the output layer to bound the reflection angles. Finally, the critic networks of both agents consist of two hidden layers, each with  256 neurons. The training hyperparameters are given in Table \ref{tab2:Hyper}.

%%%%%%%%%%%%%%%%%%%%%%%%%%%%%%%%%%%%%%%%%%%%%%%%%%%%%%%%%%%%%%%%%%%
%%%%%%%%%%%%%%--------------- RESULTS ---------------%%%%%%%%%%%%%%%
%%%%%%%%%%%%%%%%%%%%%%%%%%%%%%%%%%%%%%%%%%%%%%%%%%%%%%%%%%%%%%%%%%%%

%%%%%%%%%%%%%%%%%%%%%%%%%%%%%%%%%%%%%%%%%%%%%%%%%%%%%%%%%%%%%%
%-------------------------- Table 2 ---------------------%
%%%%%%%%%%%%%%%%%%%%%%%%%%%%%%%%%%%%%%%%%%%%%%%%%%%%%%%%%%%%
\begin{table}[t!]
\small
\centering
\caption{Algorithm Hyperparameters.}
\label{tab2:Hyper}
\begin{tabularx}{\columnwidth}{|>{\raggedright\arraybackslash}X|>{\raggedleft\arraybackslash}X|}
\hline
\textbf{Hyperparameter} & \textbf{Value} \\
\hline
Learning rate for training actor network \(\alpha_a\)  & \(1 \times 10^{-4}\) \\
\hline
Learning rate for training critic networks \(\alpha_c\)  & \(1 \times 10^{-3}\) \\
\hline
Decay factor for action exploration \(\epsilon\) & 0.9999 \\
\hline
Learning rate for soft update \(\tau_o\) & 0.001 \\
\hline
Replay buffer size \(\mathcal{D}\) & 100,000 \\
\hline
Number of experiences in the minibatch \(\mathcal{N}_b\) & 128 \\
\hline
Number of episodes & 2000 \\
\hline
Number of steps in each episode & 100 \\
\hline
Discount factor \(\gamma\) & 0.99 \\
\hline
\end{tabularx}
\end{table}

%%%%%%%%%%%%%%%%%%%%%%%%%%%%%%%%%%%%%%%%%%%%%%%%%%%%%%%%%%%
        %------------ Figure 4 -----------%
%%%%%%%%%%%%%%%%%%%%%%%%%%%%%%%%%%%%%%%%%%%%%%%%%%%%%%%%%%%
\begin{figure}[t!]
    \centering
\includegraphics[width=0.5\textwidth,height=7.2cm,width=9.5cm]{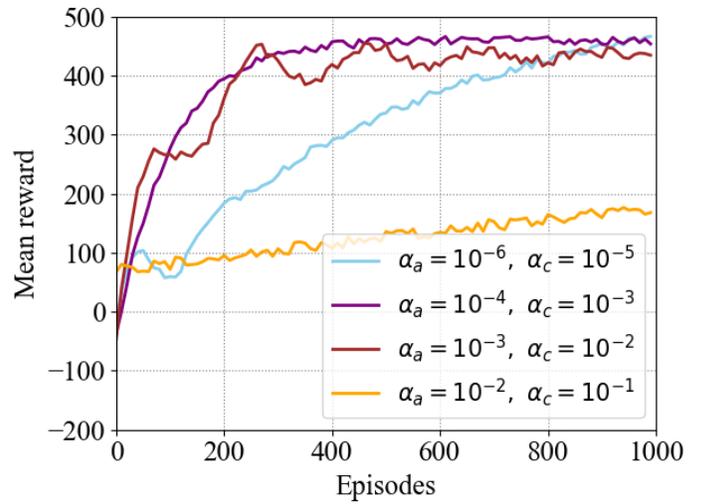}
\caption{Convergence analysis of the proposed DRL algorithm with different learning rates.} 
\vspace{-15pt}
\label{fig4}
\end{figure}
%%%%%%%%%%%%%%%%%%%%%%%%%%%%%%%%%%%%%%%%%%%%%%%%%%%%%%%%%%%%%%%%%
\subsection{Performance Analysis and Results} 
In this subsection, we evaluate the performance of the proposed DRL algorithm and compare it to various benchmark schemes under key performance metrics, including SEE, sum rate, and fairness. 

Fig. \ref{fig4} presents the impact of different learning rates, $\alpha_a$ and $\alpha_c$, for the actor and critic networks, respectively, on the agent's performance, i.e., mean reward. The results show that with a low learning rate, $(1e^{-6})$ and $(1e^{-5})$, the algorithm shows stable convergence and is more smooth. However, the low learning rate may slow the convergence of the algorithm. This means that the agent might require many more episodes to achieve a high mean reward. On the other hand, when we use high learning rates, $(1e^{-2})$ and $(1e^{-1})$, the agent suffers from training instability and increased performance variance. This is due to the fact that the neural networks are prevented from learning an effective policy or good value estimation. The figure shows that medium learning rate values, $(1e^{-4})$ and $(1e^{-3})$, demonstrate better overall performance, achieving an optimal balance between stability and convergence speed. Therefore, we set the actor and critic learning rates to $(1e^{-4})$ and $(1e^{-3})$, respectively, where the critic network maintains a higher learning rate to ensure faster value estimation, while the actor's lower rate is for stable policy updates and better convergence training in our IRS-assisted NOMA-VLC system.

%%%%%%%%%%%%%%%%%%%%%%%%%%%%%%%%%%%%%%%%%%%%%%%%%%%%%%%%%%%
%-------------------------- Figure 5 --------------------%
%%%%%%%%%%%%%%%%%%%%%%%%%%%%%%%%%%%%%%%%%%%%%%%%%%%%%%%%%%%
\begin{figure}[t]
    \centering
\includegraphics[width=0.5\textwidth,height=7cm,width=9cm]{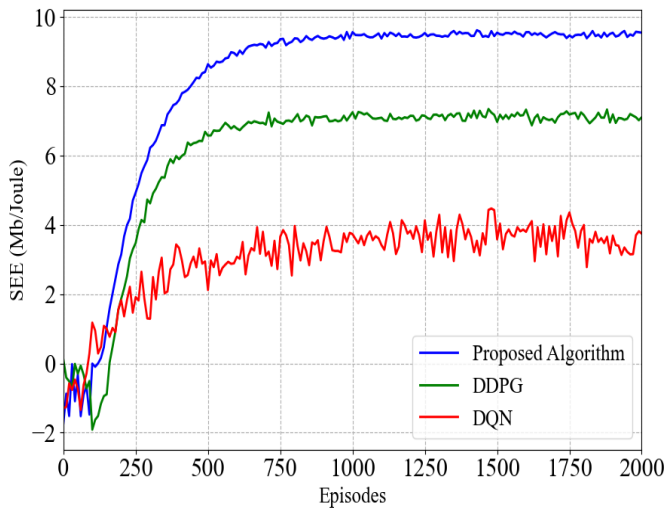}
\caption{Sum energy efficiency per episode during the training phase for different DRL algorithms.}
\vspace{-15pt}
\label{fig5}
\end{figure}

%%%%%%%%%%%%%%%%%%%%%%%%%%%%%%%%%%%%%%%%%%%%%%%%%%%%%%%%%%%
%-------------------------- Figure 6 --------------------%
%%%%%%%%%%%%%%%%%%%%%%%%%%%%%%%%%%%%%%%%%%%%%%%%%%%%%%%%%%%

\begin{figure}[t]
    \centering
    \includegraphics[width=0.5\textwidth,height=7cm,width=9cm]{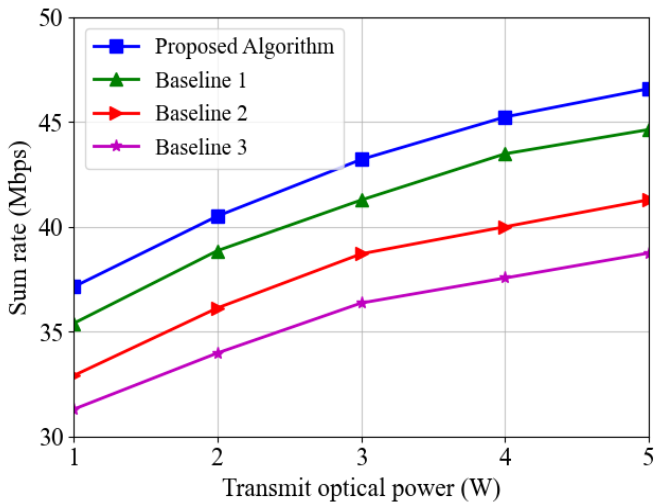}
\caption{ Sum rates versus the transmitted optical power. $M=7 \times7$.}.
\vspace{-20pt}
\label{fig:6}
\end{figure}

%%%%%%%%%%%%%%%%%%%%%%%%%%%%%%%%%%%%%%%%%%%%%%%%%%%%%%
%-------------------------- 7 ------------------%
%%%%%%%%%%%%%%%%%%%%%%%%%%%%%%%%%%%%%%%%%%%%%%%%%%%%%%
\begin{figure}[t]
    \centering
    \includegraphics[width=0.5\textwidth,height=7cm,width=9cm]{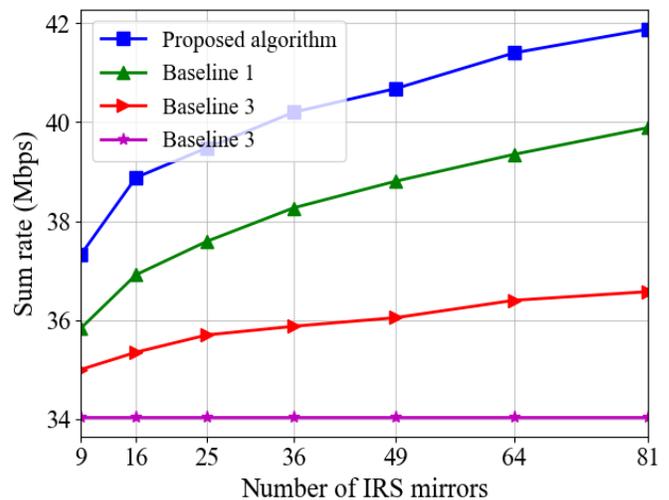}
\caption{ Sum rates versus numbers of IRS reflective mirrors $M$. Transmitted optical power, $P= 2 W$.}
\vspace{-10pt}
\label{fig7}
\end{figure}
%%%%%%%%%%%%%%%%%%%%%%%%%%%%%%%%%%%%%%%%%%%%%%%%%%%%%%%%%%%

Fig. \ref{fig5} presents the SEE performance of the proposed two-agent DRL algorithm compared to other standard DRL algorithms such as  DDPG and DQN algorithms. It can be seen that our algorithm outperforms the other two algorithms and has more stable convergence. The proposed algorithm achieves a maximum SEE of approximately 9.7 Mbits/Joule at around 500 episodes, which is better than DQL and DDPG. This convergence is attributed to its two-agent cooperation architecture, which enables more effective exploration of the joint action space through coordinated learning between the agents. Moreover, the average DDPG performance is less efficient compared to the proposed algorithm by approximately 38.5\%, and the DQN algorithm has the lowest performance with high variance. 
Note that, the proposed algorithm allows each agent to learn very sophisticated policies independently, and through CTDE learning, they remain coordinated by sharing the critic network information. While DQN and DDPG try to learn a single global value function.

Fig. \ref{fig:6} shows the sum rate versus the optical transmitted power to evaluate the performance of the proposed algorithm against other baselines introduced as follows:

\begin{itemize}
    \item  \textbf{Baseline 1:} As in Fig. \ref{fig5}, standard DDPG is implemented to solve the formulated optimization problem \textbf{P1}.
    
    \item  \textbf{Baseline 2:} In this benchmark,  a standard DDPG is considered to solve the formulated optimization problem under only the power allocation constraints, while each IRS mirror points to a specific direction. 

    \item \textbf{Baseline 3:} In this scheme, the network is remodeled without the deployment of IRS. Then, the optimization problem is reformatted under power allocation constraints and solved using a standard DDPG.
    
    \end{itemize} 
 
Fig. \ref{fig:6} shows that the sum rate increases as the transmitted optical power increases across all the schemes. Moreover, the figure shows that the proposed algorithm achieves a sum rate increase from approximately 37 Mbits/s  at 1 W to 46.5 Mbits/s at 5 W. This represents a significant improvement compared to the other baselines. The second-best performance is achieved by baseline 1, which shows a similar increase and maintains a consistent gap below our proposed algorithm, reaching approximately 38.7 Mbits/s at 2 W. Baseline 2 shows lower performance, 36 Mbits/s,  at the same transmitted power. This substantial performance gap compared to the optimized IRS orientations approaches highlights the importance of adjusting the  IRS angles in dynamic scenarios. The last baseline shows the lowest performance, highlighting the significant benefits of IRS deployment in OWC networks, achieving a 23.5\% improvement over non-IRS.

%%%%%%%%%%%%%%%%%%%%%%%%%%%%%%%%%%%%%%%%%%%%%%%%%%%%%%%%%%%
%-------------------------- Figure 8 ------------------%
%%%%%%%%%%%%%%%%%%%%%%%%%%%%%%%%%%%%%%%%%%%%%%%%%%%%%%%%%%%
\begin{figure}[t]
    \centering
    \includegraphics[width=0.5\textwidth,height=7cm,width=9cm]{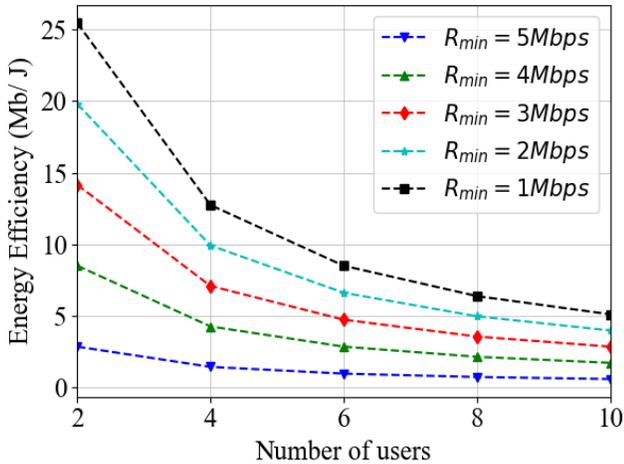}
\caption{ Energy efficiency versus numbers of users.}
\vspace{-15pt}
\label{fig8}
\end{figure}

%%%%%%%%%%%%%%%%%%%%%%%%%%%%%%%%%%%%%%%%%%%%%%%%%%%%%%%%%%%
%-------------------------- Figure 9 --------------------%
%%%%%%%%%%%%%%%%%%%%%%%%%%%%%%%%%%%%%%%%%%%%%%%%%%%%%%%%%%%
\begin{figure}[t]
    \centering
    \includegraphics[width=0.5\textwidth,height=7cm,width=9cm]{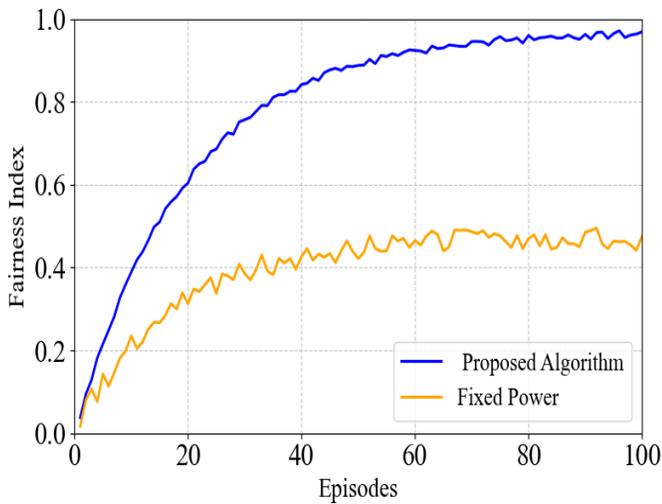}
\caption{ Fairness index over episodes.}
\vspace{-15pt}
\label{fig9}
\end{figure}

Fig. \ref{fig7} shows the impact of the IRS mirror number on the sum rates of the schemes considered. It can be seen that increasing IRS mirror elements enhances the sum rate. The proposed DRL algorithm consistently outperforms all baseline approaches considering different numbers of IRS mirrors. The proposed algorithm achieves a sum rate of 40.5 Mbps when the number of mirrors equals to $7 \times 7$. This improves the sum rate by 16.5\% over baseline 1, and 44.1\%  compared to baseline 2, in which the IRS configuration is randomly selected, and a 66.7\% enhancement over baseline 3 without IRS. However, increasing the number of IRS elements results in higher computational complexity.

Fig. \ref{fig8} shows the energy efficiency of the proposed DRL algorithm versus different numbers of users under varying minimum user rate $(R_{\min})$ requirements. It can be seen that the energy efficiency of the network decreases as the number of users increases due to the limitations of NOMA \cite{sadat2022survey}. The other reason is that the power consumption increases towards the maximum permissible power $(P_{\max})$ in the network as the number of users increases. The figure also shows that the energy efficiency decreases gradually as $R_{min}$ increases since the system consumes more power to maintain high QoS in multi-user scenarios.

%%%%%%%%%%%%%%%%%%%%%%% {Fairness Performance} %%%%%%%%%%%%%%%%%%%%%%%%%%%%%
Another essential metric to measure in our network is fairness. In Fig. \ref{fig9}, the fairness index versus a set of episodes is depicted, comparing the proposed DRL scheme to a fixed power allocation scheme. It can be seen that the fairness index rapidly converges to a value around 0.97 after only a few episodes when employing the proposed DRL. This shows the effectiveness of the proposed solution in ensuring fair distribution of power among the users. Moreover, an 80\% improvement is achieved compared to the fixed power allocation scheme, in which power allocation is predefined and fixed regardless of the network conditions. It is worth pointing out that the proposed DRL algorithm allocates the power dynamically in response to real-time updates in the network.

%%%%%%%%%%%%%%%%%%%%%%%%%%%%%%%%%%%%%%%%%%%%%%%%%%%%%%%%%%%%%%%%%%%%
%%%%%%%%%%%%%%------------- Conclusion --------------%%%%%%%%%%%%%%%
%%%%%%%%%%%%%%%%%%%%%%%%%%%%%%%%%%%%%%%%%%%%%%%%%%%%%%%%%%%%%%%%%%%%
\section{Conclusion}\label{sec:con}
In this paper, a novel two-agent DRL algorithm was proposed for real-time power allocation and IRS mirror orientation in IRS-aided NOMA-based OWC systems. We first modeled a dynamic NOMA-VLC system with IRS to enhance the coverage in scenarios of blockage and user mobility. Then, an optimization problem was formulated to jointly maximize the  SEE  and fairness. The application of traditional and deterministic optimization techniques is impractical in real time due to their high computational complexities. Therefore, the optimization problem was reformulated as MDP, and a two-agent DRL algorithm was designed based on centralized learning and decentralized execution for instantaneous solutions in highly dynamic OWC scenarios. The results show that the proposed DRL achieves a significant improvement in system performance compared to standard DRL algorithms. The results also show that the proposed algorithm achieves high performance compared to deployments without IRS and with randomly oriented IRS elements considered for the purpose of comparison.

%%%%%%%%%%%%%%%%%%%%%%%%%%%%%%%%%%%%%%%%%%%%%%%%%%%%%%%%%%%%%%%%%%%%%%
%%%%%%%%%%%%%%-------------ACKNOLEDGMENT --------------%%%%%%%%%%%%%%%
%%%%%%%%%%%%%%%%%%%%%%%%%%%%%%%%%%%%%%%%%%%%%%%%%%%%%%%%%%%%%%%%%%%%%%
\section*{Acknowledgment}

This work has been supported by the Engineering and Physical Science Research Council (EPSRC), by the INTERNET project under Grant EP/H040536/1,  by the STAR project under Grant EP/K016873/1, by the TOWS project under Grant EP/S016570/1, and by the TITAN project under Grant EP/X04047X/2. All data are provided in full in the results section of this paper.

%%%%%%%%%%%%%%%%%%%%%%%%%%%%%%%%%%%%%%%%%%%%%%%%%%%%%%%%%%%%%%%%%%%%%%%%%%%%%%%%%%%%%%%%%%%%%%%%%%%%%%%%%%%%%%

\begingroup
\footnotesize 
\bibliographystyle{IEEEtran} 
\bibliography{References} 

\begin{thebibliography}{10}
\providecommand{\url}[1]{#1}
\csname url@samestyle\endcsname
\providecommand{\newblock}{\relax}
\providecommand{\bibinfo}[2]{#2}
\providecommand{\BIBentrySTDinterwordspacing}{\spaceskip=0pt\relax}
\providecommand{\BIBentryALTinterwordstretchfactor}{4}
\providecommand{\BIBentryALTinterwordspacing}{\spaceskip=\fontdimen2\font plus
\BIBentryALTinterwordstretchfactor\fontdimen3\font minus \fontdimen4\font\relax}
\providecommand{\BIBforeignlanguage}[2]{{%
\expandafter\ifx\csname l@#1\endcsname\relax
\typeout{** WARNING: IEEEtran.bst: No hyphenation pattern has been}%
\typeout{** loaded for the language `#1'. Using the pattern for}%
\typeout{** the default language instead.}%
\else
\language=\csname l@#1\endcsname
\fi
#2}}
\providecommand{\BIBdecl}{\relax}
\BIBdecl

\bibitem{qidan2021towards}
A.~A. Qidan, T.~El-Gorashi, and J.~M. Elmirghani, ``Towards terabit lifi networking,'' \emph{arXiv preprint arXiv:2111.13784}, 2021.

\bibitem{haas2015visible}
H.~Haas, ``Visible light communication,'' in \emph{2015 Optical Fiber Communications Conference and Exhibition (OFC)}.\hskip 1em plus 0.5em minus 0.4em\relax IEEE, 2015, pp. 1--72.

\bibitem{aletri2020optimum}
O.~Z. Aletri, A.~A. Alahmadi, S.~O. Saeed, S.~H. Mohamed, T.~El-Gorashi, M.~T. Alresheedi, and J.~M. Elmirghani, ``Optimum resource allocation in optical wireless systems with energy-efficient fog and cloud architectures,'' \emph{Philosophical Transactions of the Royal Society A}, vol. 378, no. 2169, p. 20190188, 2020.

\bibitem{basar2019wireless}
E.~Basar, M.~Di~Renzo, J.~De~Rosny, M.~Debbah, M.-S. Alouini, and R.~Zhang, ``Wireless communications through reconfigurable intelligent surfaces,'' \emph{IEEE access}, vol.~7, pp. 116\,753--116\,773, 2019.

\bibitem{abdelhady2020visible}
A.~M. Abdelhady, A.~K.~S. Salem, O.~Amin, B.~Shihada, and M.-S. Alouini, ``Visible light communications via intelligent reflecting surfaces: Metasurfaces vs mirror arrays,'' \emph{IEEE Open Journal of the Communications Society}, vol.~2, pp. 1--20, 2020.

\bibitem{aboagye2021intelligent}
S.~Aboagye, T.~M. Ngatched, O.~A. Dobre, and A.~R. Ndjiongue, ``Intelligent reflecting surface-aided indoor visible light communication systems,'' \emph{IEEE Communications Letters}, vol.~25, no.~12, pp. 3913--3917, 2021.

\bibitem{kassahun2022performance}
E.~Kassahun, Z.~Hailu, K.~A. Jember, and D.~Chali, ``Performance analysis of non-orthogonal multiple access in indoor li-fi network scenarios,'' in \emph{2022 International Conference on Information and Communication Technology for Development for Africa (ICT4DA)}.\hskip 1em plus 0.5em minus 0.4em\relax IEEE, 2022, pp. 1--6.

\bibitem{marshoud2018optical}
H.~Marshoud, S.~Muhaidat, P.~C. Sofotasios, S.~Hussain, M.~A. Imran, and B.~S. Sharif, ``Optical non-orthogonal multiple access for visible light communication,'' \emph{IEEE Wireless Communications}, vol.~25, no.~2, pp. 82--88, 2018.

\bibitem{OrientationsMU}
M.~Obeed, A.~M. Salhab, M.-S. Alouini, and S.~A. Zummo, ``On optimizing vlc networks for downlink multi-user transmission: A survey,'' \emph{IEEE Communications Surveys and Tutorials}, vol.~21, no.~3, pp. 2947--2976, 2019.

\bibitem{abumarshoud2023intelligent}
H.~Abumarshoud, C.~Chen, I.~Tavakkolnia, H.~Haas, and M.~A. Imran, ``Intelligent reflecting surfaces for enhanced physical layer security in noma vlc systems,'' in \emph{ICC 2023-IEEE International Conference on Communications}.\hskip 1em plus 0.5em minus 0.4em\relax IEEE, 2023, pp. 3284--3289.

\bibitem{sun2022joint}
S.~Sun, F.~Yang, J.~Song, and Z.~Han, ``Joint resource management for intelligent reflecting surface--aided visible light communications,'' \emph{IEEE Transactions on Wireless Communications}, vol.~21, no.~8, pp. 6508--6522, 2022.

\bibitem{jiang2016machine}
C.~Jiang, H.~Zhang, Y.~Ren, Z.~Han, K.-C. Chen, and L.~Hanzo, ``Machine learning paradigms for next-generation wireless networks,'' \emph{IEEE Wireless Communications}, vol.~24, no.~2, pp. 98--105, 2016.

\bibitem{arfaoui2021invoking}
M.~A. Arfaoui, M.~D. Soltani, I.~Tavakkolnia, A.~Ghrayeb, C.~M. Assi, M.~Safari, and H.~Haas, ``Invoking deep learning for joint estimation of indoor lifi user position and orientation,'' \emph{IEEE Journal on Selected Areas in Communications}, vol.~39, no.~9, pp. 2890--2905, 2021.

\bibitem{sutton2018reinforcement}
R.~S. Sutton and A.~G. Barto, \emph{Reinforcement learning: An introduction}.\hskip 1em plus 0.5em minus 0.4em\relax MIT press, 2018.

\bibitem{elgamal2021reinforcement}
A.~S. Elgamal, O.~Z. Aletri, A.~A. Qidan, T.~E. El-Gorashi, and J.~M. Elmirghani, ``Reinforcement learning for resource allocation in steerable laser-based optical wireless systems,'' in \emph{2021 IEEE Canadian Conference on Electrical and Computer Engineering (CCECE)}.\hskip 1em plus 0.5em minus 0.4em\relax IEEE, 2021, pp. 1--6.

\bibitem{ciftler2021dqn}
B.~S. Ciftler, M.~Abdallah, A.~Alwarafy, and M.~Hamdi, ``Dqn-based multi-user power allocation for hybrid rf/vlc networks,'' in \emph{ICC 2021-IEEE International Conference on Communications}.\hskip 1em plus 0.5em minus 0.4em\relax IEEE, 2021, pp. 1--6.

\bibitem{zhao2019deep}
N.~Zhao, Y.-C. Liang, D.~Niyato, Y.~Pei, M.~Wu, and Y.~Jiang, ``Deep reinforcement learning for user association and resource allocation in heterogeneous cellular networks,'' \emph{IEEE Transactions on Wireless Communications}, vol.~18, no.~11, pp. 5141--5152, 2019.

\bibitem{Danya}
D.~A. Saifaldeen, B.~S. Ciftler, M.~M. Abdallah, and K.~A. Qaraqe, ``Drl-based irs-assisted secure visible light communications,'' \emph{IEEE Photonics Journal}, vol.~14, no.~6, pp. 1--9, 2022.

\bibitem{lillicrap2015continuous}
T.~P. Lillicrap, J.~J. Hunt, A.~Pritzel, N.~Heess, T.~Erez, Y.~Tassa, D.~Silver, and D.~Wierstra, ``Continuous control with deep reinforcement learning,'' \emph{arXiv preprint arXiv:1509.02971}, 2015.

\bibitem{sun2021intelligent}
S.~Sun, T.~Wang, F.~Yang, J.~Song, and Z.~Han, ``Intelligent reflecting surface-aided visible light communications: Potentials and challenges,'' \emph{IEEE Vehicular Technology Magazine}, vol.~17, no.~1, pp. 47--56, 2021.

\bibitem{wang2022sum}
L.~Wang, T.~Zhou, and T.~Xu, ``Sum rate maximization for multi-irs assisted downlink noma with mobile users,'' in \emph{ICC 2022-IEEE International Conference on Communications}.\hskip 1em plus 0.5em minus 0.4em\relax IEEE, 2022, pp. 3778--3783.

\bibitem{hou2020reconfigurable}
T.~Hou, Y.~Liu, Z.~Song, X.~Sun, Y.~Chen, and L.~Hanzo, ``Reconfigurable intelligent surface aided noma networks,'' \emph{IEEE Journal on Selected Areas in Communications}, vol.~38, no.~11, pp. 2575--2588, 2020.

\bibitem{cheng2021downlink}
Y.~Cheng, K.~H. Li, Y.~Liu, K.~C. Teh, and H.~V. Poor, ``Downlink and uplink intelligent reflecting surface aided networks: Noma and oma,'' \emph{IEEE Transactions on Wireless Communications}, vol.~20, no.~6, pp. 3988--4000, 2021.

\bibitem{ni2021resource}
W.~Ni, X.~Liu, Y.~Liu, H.~Tian, and Y.~Chen, ``Resource allocation for multi-cell irs-aided noma networks,'' \emph{IEEE Transactions on Wireless Communications}, vol.~20, no.~7, pp. 4253--4268, 2021.

\bibitem{abumarshoud2022intelligent}
H.~Abumarshoud, B.~Selim, M.~Tatipamula, and H.~Haas, ``Intelligent reflecting surfaces for enhanced noma-based visible light communications,'' in \emph{ICC 2022-IEEE International Conference on Communications}.\hskip 1em plus 0.5em minus 0.4em\relax IEEE, 2022, pp. 571--576.

\bibitem{liu2023sum}
Z.~Liu, F.~Yang, S.~Sun, J.~Song, and Z.~Han, ``Sum rate maximization for noma-based vlc with optical intelligent reflecting surface,'' \emph{IEEE Wireless Communications Letters}, 2023.

\bibitem{gao2021machine}
X.~Gao, Y.~Liu, X.~Liu, and L.~Song, ``Machine learning empowered resource allocation in irs aided miso-noma networks,'' \emph{IEEE Transactions on Wireless Communications}, vol.~21, no.~5, pp. 3478--3492, 2021.

\bibitem{hammadi}
A.~A. hammadi, L.~Bariah, S.~Muhaidat, M.~Al-Qutayri, P.~C. Sofotasios, and M.~Debbah, ``Deep q-learning-based resource allocation in noma visible light communications,'' \emph{IEEE Open Journal of the Communications Society}, vol.~3, pp. 2284--2297, 2022.

\bibitem{10511290}
D.~A. Saifaldeen, A.~M. Al-Baseer, B.~S. Ciftler, M.~M. Abdallah, and K.~A. Qaraqe, ``Drl-based irs-assisted secure hybrid visible light and mmwave communications,'' \emph{IEEE Open Journal of the Communications Society}, vol.~5, pp. 3007--3020, 2024.

\bibitem{zhang2024visible}
L.~Zhang, X.~Jia, N.~Tian, C.~S. Hong, and Z.~Han, ``When visible light communication meets ris: a soft actor-critic approach,'' \emph{IEEE Wireless Communications Letters}, 2024.

\bibitem{zhang2025multi}
\BIBentryALTinterwordspacing
J.~Zhang, Z.~Liu, Y.~Zhu, E.~Shi, B.~Xu, C.~Yuen, D.~Niyato, M.~Debbah, S.~Jin, B.~Ai, and X.~S. Shen, ``Multi-agent reinforcement learning in wireless distributed networks for 6g,'' \emph{arXiv preprint arXiv:2502.05812}, 2025. [Online]. Available: \url{https://arxiv.org/abs/2502.05812}
\BIBentrySTDinterwordspacing

\bibitem{bettstetter2004stochastic}
C.~Bettstetter, H.~Hartenstein, and X.~P{\'e}rez-Costa, ``Stochastic properties of the random waypoint mobility model,'' \emph{Wireless networks}, vol.~10, pp. 555--567, 2004.

\bibitem{arfaoui2022comp}
M.~A. Arfaoui, A.~Ghrayeb, C.~Assi, and M.~Qaraqe, ``Comp-assisted noma and cooperative noma in indoor vlc cellular systems,'' \emph{IEEE Transactions on Communications}, vol.~70, no.~9, pp. 6020--6034, 2022.

\bibitem{komine2004fundamental}
T.~Komine and M.~Nakagawa, ``Fundamental analysis for visible-light communication system using led lights,'' \emph{IEEE transactions on Consumer Electronics}, vol.~50, no.~1, pp. 100--107, 2004.

\bibitem{chen2017performance}
C.~Chen, W.-D. Zhong, H.~Yang, and P.~Du, ``On the performance of mimo-noma-based visible light communication systems,'' \emph{IEEE Photonics Technology Letters}, vol.~30, no.~4, pp. 307--310, 2017.

\bibitem{maraqa2023optical}
O.~Maraqa, S.~Aboagye, and T.~M. Ngatched, ``Optical star-ris-aided vlc systems: Rsma versus noma,'' \emph{IEEE Open Journal of the Communications Society}, 2023.

\bibitem{wang2013tight}
J.-B. Wang, Q.-S. Hu, J.~Wang, M.~Chen, and J.-Y. Wang, ``Tight bounds on channel capacity for dimmable visible light communications,'' \emph{Journal of Lightwave Technology}, vol.~31, no.~23, pp. 3771--3779, 2013.

\bibitem{maraqa2023optimized}
O.~Maraqa and T.~M. Ngatched, ``Optimized design of joint mirror array and liquid crystal-based ris-aided vlc systems,'' \emph{IEEE Photonics Journal}, vol.~15, no.~4, pp. 1--11, 2023.

\bibitem{aboagye2021energy}
S.~Aboagye, T.~M. Ngatched, O.~A. Dobre, and A.~G. Armada, ``Energy efficient subchannel and power allocation in cooperative vlc systems,'' \emph{IEEE Communications Letters}, vol.~25, no.~6, pp. 1935--1939, 2021.

\bibitem{jain1984quantitative}
R.~K. Jain, D.-M.~W. Chiu, W.~R. Hawe \emph{et~al.}, ``A quantitative measure of fairness and discrimination,'' \emph{Eastern Research Laboratory, Digital Equipment Corporation, Hudson, MA}, vol.~21, no.~1, 1984.

\bibitem{tahira2019optimization}
Z.~Tahira, H.~M. Asif, A.~A. Khan, S.~Baig, S.~Mumtaz, and S.~Al-Rubaye, ``Optimization of non-orthogonal multiple access based visible light communication systems,'' \emph{IEEE Communications Letters}, vol.~23, no.~8, pp. 1365--1368, 2019.

\bibitem{zhang2021multi}
K.~Zhang, Z.~Yang, and T.~Ba{\c{s}}ar, ``Multi-agent reinforcement learning: A selective overview of theories and algorithms,'' \emph{Handbook of reinforcement learning and control}, pp. 321--384, 2021.

\bibitem{dankwa2019twin}
S.~Dankwa and W.~Zheng, ``Twin-delayed ddpg: A deep reinforcement learning technique to model a continuous movement of an intelligent robot agent,'' in \emph{Proceedings of the 3rd international conference on vision, image and signal processing}, 2019, pp. 1--5.

\bibitem{jia2022policy}
Y.~Jia and X.~Y. Zhou, ``Policy gradient and actor-critic learning in continuous time and space: Theory and algorithms,'' \emph{Journal of Machine Learning Research}, vol.~23, no. 275, pp. 1--50, 2022.

\bibitem{9576103}
D.~Wang and M.~Hu, ``Deep deterministic policy gradient with compatible critic network,'' \emph{IEEE Transactions on Neural Networks and Learning Systems}, vol.~34, no.~8, pp. 4332--4344, 2023.

\bibitem{sadat2022survey}
H.~Sadat, M.~Abaza, A.~Mansour, and A.~Alfalou, ``A survey of noma for vlc systems: Research challenges and future trends,'' \emph{Sensors}, vol.~22, no.~4, p. 1395, 2022.

\end{thebibliography}
\endgroup

\end{document}